\documentclass[twoside,leqno,twocolumn]{article}
\usepackage{ltexpprt}

\usepackage{algorithm}
\usepackage[noend]{algpseudocode} % needed for pseudocode sections
\usepackage{tikz}
\usepackage{caption}
\usepackage{url}

\algrenewcommand\Return{\State \algorithmicreturn{} }

\newcommand{\pluseq}{\mathrel{+}=}
\newcommand{\minuseq}{\mathrel{-}=}

\begin{document}

\title{\Large A Chronological Edge-Driven Approach to Temporal Subgraph Isomorphism}
\author{Patrick Mackey\thanks{Pacific Northwest National Laboratory, Richland, WA, USA. \{patrick.mackey, katherine.porterfield, erin.fitzhenry, sutanay.choudhury, george.chin\}@pnnl.gov} \and Katherine Porterfield\footnotemark[1] \and Erin Fitzhenry\footnotemark[1] \and Sutanay Choudhury\footnotemark[1] \and George Chin Jr.\footnotemark[1]}
\date{}

\maketitle

\begin{abstract}
\small\baselineskip=9pt
Many real world networks are considered temporal networks, in which the chronological ordering of the edges has importance to the meaning of the data.  Performing temporal subgraph matching on such graphs requires the edges in the subgraphs to match the order of the temporal graph motif we are searching for.  Previous methods for solving this rely on the use of static subgraph matching to find potential matches first, before filtering them based on edge order to find the true temporal matches.  We present a new algorithm for temporal subgraph isomorphism that performs the subgraph matching directly on the chronologically sorted edges. By restricting our search to only the subgraphs with chronologically correct edges, we can improve the performance of the algorithm significantly.  We present experimental timing results to show significant performance improvements on publicly available datasets for a number of different temporal query graph motifs with four or more nodes.  We also demonstrate a practical example of how temporal subgraph isomorphism can produce more meaningful results than traditional static subgraph searches.
\end{abstract}

%\IEEEpeerreviewmaketitle

%\begin{IEEEkeywords}
%graph algorithms, motif analysis, subgraph isomorphism,  
%temporal networks
%\end{IEEEkeywords}

\section{Introduction}

An important area of complex systems research is that of temporal networks \cite{masuda2016guide}.  Temporal networks differ from static networks in that edges are time-stamped according to when the link occurred in the network.  One example is a cyber network, in which nodes may be IP addresses and the edges represent packets of data that transferred between them. Another common example are social networks, where the edges can represent communications or interactions that occur between people at different points in time.  Such networks are typically treated as multi-digraphs, where edges are directed and multiple edges can exist between pairs of nodes.  By including the temporal information, an analyst can hope to have a more informed understanding of the network than they would if they ignored the temporal ordering of edges.

One popular method for trying to analyze both static and temporal networks is subgraph pattern matching \cite{ullmann1976algorithm}.  Given a graph $G = (V_G, E_G)$ and a smaller graph of interest $M = (V_M, E_M)$, we attempt to find all matching subgraphs $H_1 ... H_n \in G$ that are isomorphic to $M$.  By doing so, we hope to find a set of interactions that match a particular pattern of interest.  An example might be a graph that represents a specific molecular pattern in a chemical network, or a graph that represents the behavior of a piece of malware in a cyber network.  Such subgraph matching often involves datasets with attributes on either the nodes or the edges, which can reduce the number of matches.  

In some use cases, only the total count of matching subgraphs is needed.  In this usage, the subgraph patterns we seek to find are typically referred to as \textit{network motifs}, and the total count of each pattern is used as a way of characterizing the network as a whole \cite{milo2002network}.  It is also possible to use the count of matching subgraphs as a way of characterizing individual nodes in a network, based on the number of matching subgraphs each node lies on.  Network motif analysis has found a wide range of uses including analyzing biological networks \cite{shen2002network}, social networks \cite{faust2002comparing} and computational graphs \cite{valverde2005network}.

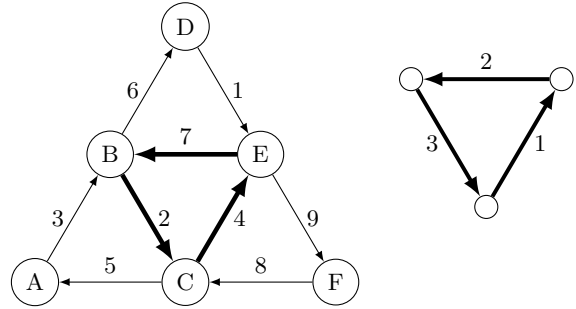
\begin{figure}
\begin{tikzpicture}
\tikzset{>=latex}
\small
\node[draw, circle] (A) at (0,0) {A};
\node[draw, circle] (B) at (1,1.7) {B};
\node[draw, circle] (C) at (2,0) {C};
\node[draw, circle] (D) at (2,3.4) {D};
\node[draw, circle] (E) at (3,1.7) {E};
\node[draw, circle] (F) at (4,0) {F};
\draw[->] (A) -- (B) node[pos=.5, left]{3};
\draw[->] (C) -- (A) node[pos=.5, above]{5};
\draw[->,line width=1.75pt] (B) -- (C) node[pos=.5, right]{2};
\draw[->,line width=1.75pt] (C) -- (E) node[pos=.5, right]{4};
\draw[->,line width=1.75pt] (E) -- (B) node[pos=.5, above]{7};
\draw[->] (D) -- (E) node[pos=.5, right]{1};
\draw[->] (B) -- (D) node[pos=.5, left]{6};
\draw[->] (F) -- (C) node[pos=.5, above]{8};
\draw[->] (E) -- (F) node[pos=.5, right]{9};
\node[draw, circle] (0) at (6,1){};
\node[draw, circle] (1) at (7,2.7){};
\node[draw, circle] (2) at (5,2.7){};
\draw[->,line width=1.5pt] (0) -- (1) node[pos=.5, right]{1};
\draw[->,line width=1.5pt] (1) -- (2) node[pos=.5, above]{2};
\draw[->,line width=1.5pt] (2) -- (0) node[pos=.5, left]{3};
\end{tikzpicture}
\caption{The temporal graph on the left has four directed 3-cycles, but only one (B,C,E) matches the sequential ordering of the temporal motif on the right.}
\end{figure}

The subgraph isomorphism problem has long been known to be NP-complete \cite{cook1971complexity}, however it can often be computed relatively efficiently in practice.  The exponential performance of the subgraph matching algorithm is usually only exponential in regards to the number of nodes or edges in $M$.  This means searching for a small subgraph can actually be quite efficient in many cases, particularly if node and edge attributes are used to restrict the number of potential matches.

While a number of techniques exist for finding matches in static networks, most do not take the temporal order of the edges into account when finding instances of $M$ in $G$.  However, the ordering of these edges be essential to the meaning of the query graph or motif.  A simple example is that of a sequential cycle, in which each edge in the cycle occurs before the one following it.  If time is not taken into account, non-sequential cycles may be found instead, which could have completely different meaning given the use case.  For an example see Figure 1, in which we are looking for a sequentially ordered 3-cycle in our graph.  While a number of 3-cycles exist, only one matches this ordering criteria correctly.

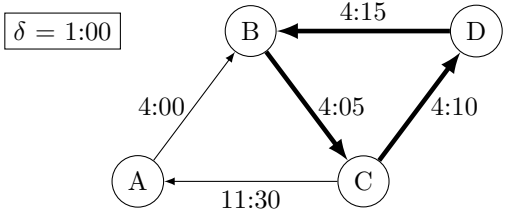
\begin{figure}
\begin{tikzpicture}
\tikzset{>=latex}
\node[draw, circle] (A) at (0,0) {A};
\node[draw, circle] (B) at (1.5,2) {B};
\node[draw, circle] (C) at (3,0) {C};
\node[draw, circle] (D) at (4.5,2) {D};
\draw[->] (A) -- (B) node[pos=.5, left]{4:00};
\draw[->,line width=1.75pt] (B) -- (C) node[pos=.5, right]{4:05};
\draw[->] (C) -- (A) node[pos=.5, below]{11:30};
\draw[->,line width=1.75pt] (C) -- (D) node[pos=.5, right]{4:10};
\draw[->,line width=1.75pt] (D) -- (B) node[pos=.5, above]{4:15};
\node[draw, rectangle] (delta) at (-1,2) {$\delta$ = 1:00}; %(6,1)
\end{tikzpicture}
\caption{This temporal graph has two sequential 3-cycles, but only one that meets our restriction $\delta$ = 1:00 (B,C,D).  The 3-cycle (A,B,C) is also sequential, but extends beyond the 1:00 time period specified by the $\delta$-temporal motif.}
\end{figure}

In addition to a simple temporal ordering restriction, it can also be useful to add a restriction on the length of time permitted between edges in the subgraph.  This is referred to as a $\delta$-temporal motif \cite{paranjape2017motifs}.  In a $\delta$-temporal motif approach, all edges in the matching subgraph must occur within a period of $\delta$ time units.  This added restriction prevents us from including edges that may have occurred long after the initial edges in our subgraph.  This can improve both the usefulness of the results as well as the performance time required to find the matching subgraphs.  See Figure 2 as an example of a sequential 3-cycle, with $\delta = 1$ hour.  In this case the cycle between A, B and C would not be included, despite having the correct ordering, due to the edge between C and A occurring several hours later.

In this paper we present a new algorithm for finding matching subgraphs in a temporal network that inherently takes the ordering of the edges into account.  We do this through a chronologically sorted edge-driven approach that only searches through edges that occur in the correct order.  By doing so, we can have a significantly more efficient approach for general temporal subgraph matching than was previously possible.

\section{Related Work}

Subgraph matching algorithms have existed since at least the time of the publication of Ullmann's algorithm in 1976 \cite{ullmann1976algorithm}.  Since then, many modifications have been made to this approach to improve its performance.  One of the most commonly used is known as the VF2 algorithm \cite{cordella2004sub}, itself an extension of the VF algorithm \cite{cordella1999performance} (named for two of its authors, Vento and Foggia).  While this approach can be used to find patterns in static networks, by default it ignores the chronological ordering of edges, and only expects them to match a particular topological configuration.  While it may be possible to add edge restrictions to enforce some temporal requirements in how the subgraphs are matched, it is not a natural part of the algorithm.

Motif analysis on temporal networks is a much more recent area of research. Kovanen et al. provide an algorithm for finding temporal subgraphs, but have the added restriction that edges in the motif must represent consecutive events for the node  \cite{kovanen2011temporal}.  While this allows for fast subgraph counting, it does not allow subgraph matching on graphs that have other edges occurring in between the edge events of a motif.  Another approach was published by Gurukar et al. \cite{gurukar2015commit}.  The authors present a heuristic for counting temporal motifs, but provide only an estimated count with their approach, and do not exhaustively find all exact temporal subgraph matches. 

Another closely related work is that of Redmond and Cunningham \cite{redmond2013temporal} \cite{redmond2016subgraph}.  Their definition of temporal subgraph isomorphism has some similarities to our own, but also has some distinct differences.  The most significant difference is that the query graphs do not have their edges time-stamped.  Instead, a subgraph is considered to be a temporal match if all of the edges occur within a particular $\delta$ and incoming edges occur before outgoing edges.  Since their query graphs do not have any particular ordering to their edges (beyond that incoming edges must occur first), there is no way to directly compare the results of their algorithm to our own.

Closer in spirit to our work is the recent publication by Parnajape, Benson and Leskovec \cite{paranjape2017motifs}.  Their research uses an equivalent definition of temporal subgraph matching to our own.  Their algorithmic approach, however, is quite different.  When it comes to the general subgraph matching use case, their approach is to first perform a static subgraph match to find potential temporal matches.  After these potential matches are found, a separate algorithm is used to find all temporal subgraph matches from these potential matches.  While the running time of this secondary algorithm is linear, it requires all potential matches to be found first.  This means that in some cases significantly more subgraphs will be initially found than needed, slowing down the overall performance of the algorithm.  In theory, this difference could be exponential, requiring exponentially more time than an approach that finds only the true temporal matches would.  A particularly bad example would be looking for a $\delta$-temporal motif in a complete graph, where only one $\delta$-temporal match existed.  Without any additional restrictions, an exponential number of subgraphs would be returned by the intial static subgraph matching phase, despite there be only one actual temporal subgraph match.

In addition to their general temporal subgraph matching algorithm, the authors also present a faster algorithms for counting 2-node, 3-edge star and triangle temporal motifs \cite{paranjape2017motifs}.  These approaches are significantly faster than their approach for general temporal subgraph matching, but have the downside of working for only a particular set of graph patterns.  While these differ significantly from both our approach and earlier static approaches, it is similar to ours in regards of being chronologically edge-driven, with the edges iterated over in chronological order to find the matching subgraphs.  Their timing experiments show them to be very efficient at finding 2-node, 3-edge star and triangle temporal motifs.  While it is common to perform motif analysis on small graphs such as these, there are many known cases where larger motifs are needed or useful \cite{valverde2005network, yeger2004network, zhang2005motifs}.  The need to support efficient temporal subgraph matching on larger query graphs was a main factor driving our own algorithm research.

\section{Example Application}

To give an example of the usefulness of temporal graph isomorphism, we will demonstrate how a targeted temporal subgraph query can produce practical results from a publicly available dataset, and how the results vary significantly if the temporal ordering is ignored.

The data used was collected from the Carnegie Mellon CERT Insider Threat Tools dataset, a synthetically generated insider cyber test dataset \cite{glasser2013bridging}.  The data represents a fictitious company, with the nodes representing the employees, their computers, files, removable media and websites they visited.  Edges represent actions connecting these nodes together (e.g., download a file, log in to a PC, etc.).  For our analysis, we have selected data taking place in the month of April 2011, which represents 315K nodes with 21.8M temporal edges.

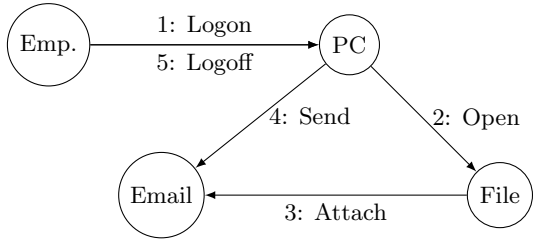
\begin{figure}
\begin{tikzpicture}
\tikzset{>=latex}
%\large
%\node at (0,6) {$M_1$};
\small
\node[draw, circle] (Person) at (1,3) {Emp.};
\node[draw, circle] (PC) at (5,3) {PC};
\node[draw, circle] (File) at (7,1) {File};
\node[draw, circle] (Email) at (2.5,1) {Email};
\draw[->] (Person) -- (PC) node[pos=.5, above]{1: Logon};
\draw[->] (PC) -- (File) node[pos=.5, right]{2: Open};
\draw[->] (File) -- (Email) node[pos=.5, below]{3: Attach};
\draw[->] (PC) -- (Email) node[pos=.5, right]{4: Send};
\draw[->] (Person) -- (PC) node[pos=.5, below]{5: Logoff};
\end{tikzpicture}
\caption{The temporal query graph used in our CERT use case example.}
\end{figure}

The particular insider threat we have focused on is the case where an employee is repeatedly stealing information from their co-workers by logging into their machines, opening files, and sending them to his home e-mail address.  A motif that may be indicative of such a threat is shown in Fig 3.  While not all actions matching this query would necessarily indicate malicious behavior, observing an employee with an unusually high rate of matching subgraphs may indicate closer observation may be needed.  For our use case, we have selected a $\delta$ = 1 hour, as we are expecting this pattern to represent actions that occur in a relatively short period of time.  When searching for this temporal query graph in our dataset, we find the exact employee we were expecting (CDE1846), who was in fact the perpetrator in the dataset.  

We also compared how searching for this same motif with an unordered static subgraph isomorphism algorithm would effect the results.  The results of our experiments can be seen in Table 1.  The first column represents the subgraph isomorphism technique being used.  The second column has the value of $\delta$ (where appropriate).  The third column lists the number of employee nodes that were found on matching subgraphs.  The fourth column tells the rank of the perpetrator node in our search.  A value of 1 means the perpetrator was found on the largest number of matching subgraphs.

\begin{table}
\begin{footnotesize}
\begin{tabular}{|llcc|}
\hline
  \textbf{Isomorphism} & \textbf{$\delta$} & \textbf{\# Matches} & \textbf{Rank} \\
Temporal & 1 hr & 1 & 1 \\
Temporal & 2 hr & 8 & 1 \\
Temporal & 4 hr & 68 & 1 \\
Temporal & 8 hr & 298 & 1 \\
Temporal & 1 day & 600 & 4 \\
Temporal & 2 days & 721 & 47 \\
Temporal (Alt. Order) & 1 hr & 2 & 1 \\
Temporal (Alt. Order) & 2 hr & 28 & 1 \\
Temporal (Alt. Order) & 4 hr & 343 & 1 \\
Temporal (Alt. Order) & 8 hr & 1,053 & 7 \\
Temporal (Alt. Order) & 1 day & 2,247 & 24 \\
Temporal (Alt. Order) & 2 days & 2,248 & 132 \\
Static & NA & 2,297 & 410 \\
\hline
\end{tabular}
\end{footnotesize}
\caption{Results from subgraph queries for our CERT use case example. \# Matches represents the number of employees found on matching subgraphs.  Rank represents how high the actual perpetrator was in the ranking of nodes (with 1 being the highest).}
\end{table}

While our targeted temporal search gave us the exact correct answer, a static search gave back a much larger number of employees lying on matching subgraphs.  A total of 2,297 matching nodes were found, with the perpetrator having a rank of only 410 in the list.  There are actually a number of reasons why temporal order is so important in this use case.  One important factor is the use of the $\delta$ value. By restricting the time to one hour, we are describing a particular set of actions: one that occurs in a relatively short amount of time.  If we make this value too high, we can get inferior results, even with the correctly temporal ordering.  When this same pattern occurs over a period of several hours, it may represent a normal employee going about his normal daily business, instead of the pattern we are actually looking for, which is someone logging into multiple PCs quickly to pilfer files from their co-workers.  As can be seen in Table 1, as the value of $\delta$ increases, we get increasingly more employees found in matching subgraphs, despite the fact these employees were innocent.  When $\delta$ increases to the size of 1 day, the ranking of the perpetrator drops as well.

Another important factor is that there is a different meaning depending on the order of these events/edges.  A user who is opening a file after sending it suggests he is already familiar with the contents.  A user who opens it first suggests someone who is unsure of the contents, and may be looking for something in particular before sending it.  While the difference between these actions may be subtle, they do represent actual differences in behavior.  As an example, if we flip the order of these events, we get significantly worse results, as can be seen in the rows marked Temporal (Alt. Order).  Not only is there an increasing number of matching employees, but we also begin to see the actual perpetrator drop in rank as $\delta$ increases.

\section{Approach}

While our approach could be modified to do simple subgraph counting, our desire was to return a set of all instances of subgraphs $H_1 ... H_n \in G$ that were isomorphic to $M$.  We also wanted an algorithm that was not limited to any particular size or structure of query graph or motif.  The heart of the algorithm is described in the function \textsc{TemporalMatch} in \textbf{Algorithm 1} as well as the related subroutine \textsc{FindNextMatch} listed in \textbf{Algorithm 2}.  Some important bookkeeping is necessary to make sure the nodes and edges of the graph are being matched correctly:

\begin{algorithm}
\caption{Algorithm for chronologically sorted edge-driven temporal subgraph matching.  Exhaustively finds all subgraphs isomorphic to our motif, including the chronological ordering of their edges.}
\label{TemporalMatch}\footnotesize
\begin{algorithmic}[1]
%\State \textbf{Inputs:} 
%\State $G = (V_G,E_G)$, the temporal multi-digraph we are searching on, with edges $E_G$ sorted chronologically.
%\State $M = (V_M,E_M)$, the temporal multi-digraph motif we are looking for, with edges $E_M$ sorted chronologically.
%\State $\delta$, the maximum time period permitted between edges in our matching subgraph.
%\State \textbf{Output:} Set of all matching subgraphs $H_1 ... H_n \in G$ that are isomorphic to $M$.
\Function{TemporalMatch}{$G,M,\delta$}
	\State \textit{--- Initialize necessary variables ---}
	\For{$v_G \in V_G$}
		\State edgeCount$[v_G] \gets 0$
		\State map$_{GM}[v_G] \gets -1$
	\EndFor
	\For{$v_M \in V_M$}
		\State map$_{MG}[v_M] \gets -1$
	\EndFor
	\State results $\gets \{\}$
	\State eStack $\gets \{\}$
	\State $e_G \gets 0$
	\State $e_M \gets 0$
	\State $t' \gets \infty$
	\State \textit{--- Loop until all matching subgraphs are found ---}
	\While{true}
		\State $e_G \gets$ \textsc{FindNextMatch}($G,e_M,e_G,$map$_{MG}$,map$_{GM},t'$)
		\If{$e_G < |E_G|$} % and time($e_G$) - time(eStack.top()) $\leq \delta$}
			\State \textit{--- Test if all edges in $M$ are matched ---}
			\If{$e_M = |E_M| - 1$}
				\State Create graph $H$ from edges in eStack
				\State Add $H$ to results
			\Else				
				\State $(u_G,v_G) \gets E_G[e_G]$
				\State $(u_M,v_M) \gets E_M[e_M]$
				%\State \textit{--- Map nodes from $G$ to $M$ ---}
				\State map$_{GM}[u_G] \gets u_M$				
				\State map$_{GM}[v_G] \gets v_M$	
				\State map$_{MG}[u_M] \gets u_G$				
				\State map$_{MG}[v_M] \gets v_G$	
				%\State \textit{--- Increment number of searched edges ---}
				\State edgeCount$[u_G] \pluseq 1$
				\State edgeCount$[v_G] \pluseq 1$
				\If{eStack.empty()}
					\State $t' \gets$ time$(e_G) + \delta$
				\EndIf
				\State eStack.push($e_G$)
				\State $e_M \pluseq 1$
			\EndIf			
		\EndIf 
		\State $e_G += 1$		
		\State \textit{--- Backup or quit if we run out of edges ---}
		\While{$e_G \geq |E_G|$ or time$(e_G)> t'$}		
			\If{eStack is not empty}
				%\State \textsc{BackupSearch}()
				\State $e_G \gets$ eStack.pop() $+ 1$
				\If{eStack.empty()}
					\State $t' \gets \infty$
				\EndIf
				\State edgeCount$[u_G] \minuseq 1$
				\State edgeCount$[v_G] \minuseq 1$
				\State \textit{--- Unassign nodes if needed ---}
				\If{edgeCount$[u_G] = 0$}		
					\State $u_M \gets$ map$_{GM}[u_G]$
					\State map$_{MG}[u_M] \gets -1$
					\State map$_{GM}[u_G] \gets -1$
				\EndIf
				\If{edgeCount$[v_G] = 0$}		
					\State $v_M \gets$ map$_{GM}[v_G]$
					\State map$_{MG}[v_M] \gets -1$
					\State map$_{GM}[v_G] \gets -1$
				\EndIf
				\State $e_M \minuseq 1$
			\Else
				\State \textbf{return} results
			\EndIf
		\EndWhile
	\EndWhile
\EndFunction
\end{algorithmic}
\end{algorithm}

\begin{itemize}
	\item $E_G$: A list of edges $\in G$, sorted chronologically by time stamp.
	\item $E_M$: A list of edges $\in M$, sorted chronologically by time stamp.
	\item map$_{GM}$: A mapping of nodes in $G$ to nodes in $M$.  In practice, this is an array of length $|V_G|$.  If a node is not assigned to node in $M$, it will have a value of -1.
	\item map$_{MG}$: A mapping of nodes in $M$ to nodes in $G$.  In practice, this is an array of length $|V_M|$.  If a node is not yet assigned to node in $G$, it will have a value of -1.
	\item edgeCount: A mapping of nodes in $G$ to the number of adjacent edges that have been mapped to them for our motif $M$.  When this value drops to zero, that vertex is free to be reassigned to another node in $M$ if necessary.
	\item eStack: A stack used to keep track of what edges in $G$ are currently mapped to edges in $M$.  They are always mapped in chronological order, so that the first one will map to the first edge in $M$.
	\item $t'$: The latest time an edge can have in our matching subgraph, given the time for the first edge and $\delta$.  This is initially infinity until the first edge is added.  It gets reset when eStack is empty.
	\item results: A set of matching subgraphs to return at the completion of the algorithm.
\end{itemize}

\begin{algorithm}
\caption{Subroutine for finding the next matching temporal edge that matches edge $e_M$ in our motif $M$. }
\label{FindNextMatch}\footnotesize
\begin{algorithmic}[1]
%\State \textbf{Inputs:}
%\State $G = (V_G,E_G)$, the temporal multi-digraph we are searching on, with edges $E_G$ sorted chronologically.
%\State $e_M$, edge index in $E_M \in M$, that we are trying to match.
%\State $e_G$, edge index in $E_G \in G$, to begin our search for a matching edge.
%\State map$_{MG}$, the current mapping of nodes in $M$ to to nodes in $G$.
%\State map$_{GM}$, the current mapping of nodes in $G$ to to nodes in $M$.
%\State $t'$, the latest time allowed, given our first edge and $\delta$.
%\State \textbf{Output:} An edge index in $E_G$ that matches the criteria of $e_M$.  If no suitable matche is found, returns the size of $E_G$.
\Function{FindNextMatch}{$G,e_M,e_G,$map$_{MG}$,map$_{GM},t'$}
	\State $(u_M,v_M) \gets E_M[e_M]$
	\State $u_G \gets$ map$_{MG}[u_M]$
	\State $v_G \gets$ map$_{MG}[v_M]$
	\State \textit{--- Determine the potential edges to try ---}
	\State $S = E$
	\If{$u_G \geq 0$ and $v_G \geq 0$}
		\State $S \gets$ all edges $e$ between $G[u_G][v_G]$ where $e \geq e_G$ and time$(e_G) \leq t'$
	\ElsIf{$u_G \geq 0$}
		\State $S \gets$ all edges $e$ emanating from $u_G$ where $e \geq e_G$ and time$(e_G) \leq t'$
	\ElsIf{$v_G \geq 0$}
		\State $S \gets$ all edges $e$ emanating from $v_G$ where $e \geq e_G$ and time$(e_G) \leq t'$
	\EndIf
	\State \textit{--- Try each edge until a match is made ---}
	\For{$e_G' \in S$}
		\State $(u_G',v_G') \gets E_G[e_G']$
		\State \textit{--- The mapping must match, or be unassigned ---}
		\If{$u_G = u_G'$ or ($u_G < 0$ and map$_{GM}[u_G'] < 0$)}
			\If{$v_G = v_G'$ or ($v_G < 0$ and map$_{GM}[v_G'] < 0$)}
				\If{attribute match criteria are met}
					\State \textbf{return} $e_G'$
				\EndIf
			\EndIf
		\EndIf
	\EndFor
	\State \textit{--- Only reached if no suitable match found ---}
	\State \textbf{return} $|E_G|$
\EndFunction
\end{algorithmic}
\end{algorithm}

The algorithm works by iterating over each edge index $e_M$ in $E_M$ in chronological order.  Before the next edge in $M$ can be reached, a matching edge $e_G$ in $E_G$ must be found.  (Note: $e_M$ and $e_G$ are integer indices into the sorted list of edges $E_M$ and $E_G$, and not actual edge structures).  If an appropriate $e_G$ is found, we map the source and destination nodes accordingly, and continue on to find a match for the next edge in $M$.  Once the remaining edges in $M$ are accounted for, we can add that subgraph to our results list, and continue on to find an alternative edge $e_G$ in $G$ that can also map to the last edge $e_M$.  If we iterate over all remaining edges before all edges in $E_M$ are accounted for, we must pop the last matched edge from our stack, and try again from that point in the list of edges.

To determine which edge in $E_G$ we can try next, we have the \textsc{FindNextMatch} subroutine specified in \textbf{Algorithm 2}.  This takes into account whether or not either the source or destination node have already been mapped, as well as the sequential ordering of the most recently mapped edge.  If these are not taken into account, the performance of the algorithm may drastically slow down.  Other attribute based approaches may also be used to help improve this narrowing process as well.

\subsection{Time Complexity}

A worst case computational time complexity can be calculated to be $O(|E_G| ^{|E_M|})$.  This would occur when each edge in $E_G$ could be matched to each edge in $E_M$, and the value of $\delta$ exceeds that of the time range for $E_G$.  An example would be a complete graph with an exponential number of subgraphs isomorphic to the query graph. For every edge in $E_M$ we could expect it to match to every edge in $E_G$ subsequent in time to the selected edge.  Even with the temporal restriction that the edges must be subsequent to each other, we can still have an exponential asymptotic complexity, since the sum of a linearly decreasing series is still exponential (e.g., $\sum\limits_{i=0}^{n} n-i = \frac{n(n+1)}{2} = O(n^2)$).

In practice, we can expect the runtime to be typically less than this.  If the number of edges expected within $\delta$ is a constant $k$, we can calculate the time complexity to be $O(|E_G| \cdot k^{|E_M|-1})$.  The proof for this is straight forward: In a worst case scenario (where each edge in $E_G$ is a match for every edge in $E_M$) at most $k$ edges need to be visited for every edge in $E_M$.  In real-world scenarios, we can actually expect performance to be even better than this, since in most situations not every edge should be a match.  This also helps to shed light on how the value of $\delta$ can have such a large impact in the overall performance time.

In comparison to existing approaches, using VF2 to perform static subgraph matching has a worst case computational time complexity of $O(|V_G|! \cdot |V_G|)$ and a best case of $O(|V_G|^2)$ \cite{cordella2004sub}.  To perform the additional steps needed to perform the general temporal subgraph matching of Paranjape et al., there is an additional time of $O(|H|^{|E_M|} \cdot |S'|)$, where $H$ is the static subgraph induced by $M$ and $S'$ is the temporal edges between pairs of nodes in $H$ \cite{paranjape2017motifs}.  As the values of $|H|$ and $|E_M|$ tend to be very small, the authors consider this value to be essentially a constant, and consider the performance of this secondary step linear in regards to $|S'|$.  As long as the number of potential temporal edge matches ($|S'|$) does not exceed $|V_G|^2$, we can consider the overall time complexity of the approach of Paranjape et al. to be asymptotically equivalent to VF2.

\section{Timing Experiments}

Determining which method should work faster analytically is a challenge, as the performance of both approaches is exponential in a worst case scenario, and many factors can come into play in determining the exact time complexity.  Instead, we have opted to do a number of timing experiments comparing the performance time of our algorithm against that of VF2. As this is actually just the first step in the previously published approach, in practice the actual time required should be longer.  However, we believe this should give us a decent baseline to compare our performance against, as the secondary step of Paranjape et al. is performed in linear time, and the implementation of their general temporal subgraph isomorphism algorithm is currently unavailable to us.

\setlength{\tabcolsep}{0.35em} % for the horizontal padding
\begin{table}
\begin{footnotesize}
\begin{tabular}{|ccccc|}
\hline
  \textbf{Graph} & \textbf{Nodes} & \textbf{Static Edges} & \textbf{Edges} & \textbf{Time Span} \\
  CollegeMsg & 1.9K & 20.3K & 60K & 193 days\\
  Email-Eu & 986 & 2.5K & 332K & 2.2 years \\
  %Wiki-Vote & 7.12K & 104K & 104K & 3.5 years \\
  MathOverflow & 24.8K & 228K & 507K & 6.5 years \\  
  Enron & 82.2K & 322K & 1.15M & 4 years \\
\hline
\end{tabular}
\end{footnotesize}
\caption{Graph statistics for experimental data sets.  Static edges are reduced in size by merging parallel temporal edges together.}
\end{table}

For our experiments we have selected four temporal network datasets from the SNAP graph library \cite{snapnets} and Koblenz Network Collection (KONECT) \cite{kunegis2013konect}. Information about each of the datasets can be seen in Table 2.  \textbf{CollegeMsg} represents a set of private messages between users of an online social network at University of California, Irvine \cite{panzarasa2009patterns}.  \textbf{Email-Eu} represents a collection of emails between members of an anonymous European research institution \cite{paranjape2017motifs}.  \textbf{MathOverflow} represents comments on a popular online messageboard on mathematics, where an edge represents a comment between users \cite{paranjape2017motifs}.  \textbf{Enron} contains edges representing the email traffic between employees of the Enron corporation between 1999 and 2003 \cite{kunegis2013konect}.  Each networks has its own unique topological characteristics, which can be seen to some extent by the varying results from the temporal motif counting in Table 3.

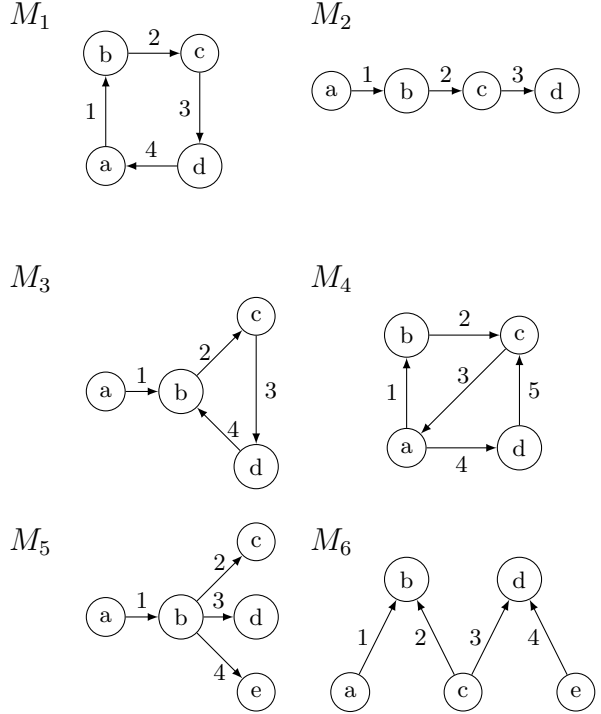
\begin{figure}
\begin{tikzpicture}
\tikzset{>=latex}
\large
\node at (0,6) {$M_1$};
\node at (4,6) {$M_2$};
\node at (0,2.5) {$M_3$};
\node at (4,2.5) {$M_4$};
\node at (0,-1) {$M_5$};
\node at (4,-1) {$M_6$};
\small
\node[draw, circle] (A1) at (1,4) {a};
\node[draw, circle] (B1) at (1,5.5) {b};
\node[draw, circle] (C1) at (2.25,5.5) {c};
\node[draw, circle] (D1) at (2.25,4) {d};
\node[draw, circle] (A2) at (4,5) {a};
\node[draw, circle] (B2) at (5,5) {b};
\node[draw, circle] (C2) at (6,5) {c};
\node[draw, circle] (D2) at (7,5) {d};
\node[draw, circle] (A3) at (1,1) {a};
\node[draw, circle] (B3) at (2,1) {b};
\node[draw, circle] (C3) at (3,2) {c};
\node[draw, circle] (D3) at (3,0) {d};
\node[draw, circle] (A4) at (5,.25) {a};
\node[draw, circle] (B4) at (5,1.75) {b};
\node[draw, circle] (C4) at (6.5,1.75) {c};
\node[draw, circle] (D4) at (6.5,.25) {d};
\node[draw, circle] (A5) at (1,-2) {a};
\node[draw, circle] (B5) at (2,-2) {b};
\node[draw, circle] (C5) at (3,-1) {c};
\node[draw, circle] (D5) at (3,-2) {d};
\node[draw, circle] (E5) at (3,-3) {e};
\node[draw, circle] (A6) at (4.25,-3) {a};
\node[draw, circle] (B6) at (5,-1.5) {b};
\node[draw, circle] (C6) at (5.75,-3) {c};
\node[draw, circle] (D6) at (6.5,-1.5) {d};
\node[draw, circle] (E6) at (7.25,-3) {e};
\draw[->] (A1) -- (B1) node[pos=.5, left]{1};
\draw[->] (B1) -- (C1) node[pos=.5, above]{2};
\draw[->] (C1) -- (D1) node[pos=.5, left]{3};
\draw[->] (D1) -- (A1) node[pos=.5, above]{4};
\draw[->] (A2) -- (B2) node[pos=.5, above]{1};
\draw[->] (B2) -- (C2) node[pos=.5, above]{2};
\draw[->] (C2) -- (D2) node[pos=.5, above]{3};
\draw[->] (A3) -- (B3) node[pos=.5, above]{1};
\draw[->] (B3) -- (C3) node[pos=.5, left]{2};
\draw[->] (C3) -- (D3) node[pos=.5, right]{3};
\draw[->] (D3) -- (B3) node[pos=.5, right]{4};
\draw[->] (A4) -- (B4) node[pos=.5, left]{1};
\draw[->] (B4) -- (C4) node[pos=.5, above]{2};
\draw[->] (C4) -- (A4) node[pos=.5, above]{3};
\draw[->] (A4) -- (D4) node[pos=.5, below]{4};
\draw[->] (D4) -- (C4) node[pos=.5, right]{5};
\draw[->] (A5) -- (B5) node[pos=.5, above]{1};
\draw[->] (B5) -- (C5) node[pos=.5, above]{2};
\draw[->] (B5) -- (D5) node[pos=.5, above]{3};
\draw[->] (B5) -- (E5) node[pos=.5, below]{4};
\draw[->] (A6) -- (B6) node[pos=.5, left]{1};
\draw[->] (C6) -- (B6) node[pos=.5, left]{2};
\draw[->] (C6) -- (D6) node[pos=.5, left]{3};
\draw[->] (E6) -- (D6) node[pos=.5, left]{4};
\end{tikzpicture}
\caption{The temporal motifs for our experimental analysis.}
\end{figure}

\begin{table*}[t]
\begin{small}
\begin{tabular}{|lc|cc|ccc|ccc|}
\hline
& & \textbf{Static} & & \textbf{$\delta$ = 1 hour} & & & \textbf{$\delta$ = 1 day} & & \\
  \textbf{Graph} & \textbf{Motif} & \textbf{Subgraphs} & \textbf{Time} & \textbf{Subgraphs} & \textbf{Time} & \textbf{Speed Up} & \textbf{Subgraphs} & \textbf{Time} & \textbf{Speed Up} \\
\hline
CollegeMSG & $M_1$ & 876 & 18.8 & 3.46K & 0.546 & 34.5x & 63.2K & 6.24 & 3.01x \\
CollegeMSG & $M_2$ & 238K & 43.2 & 75.4K & 2.04 & 21.1x & 1.46M & 34.3 & 1.26x\\
CollegeMSG & $M_3$ & 1.57K & 19.3 & 1.06K & 0.499 & 38.6x & 53.5K & 6.16 & 3.12x \\
CollegeMSG & $M_4$ & 48 & 2.18 & 0 & 0.296 & 7.38x & 173K & 1.28 & 1.71x \\
CollegeMSG & $M_5$ & 202M & 9.88K & 1.28M & 35.1 & 281x & 49.7M & 1.34K & 7.36x \\
CollegeMSG & $M_6$ & 10.9M & 2.02K & 227K & 6.74 & 300x & 16.2M & 435 & 4.66x \\
\hline
Email-Eu & $M_1$ & 3.96K & 33.5 & 258 & 1.14 & 29.4x & 129K & 12.9 & 2.60x \\
Email-Eu & $M_2$ & 367K & 91.4 & 30.1K & 1.80 & 50.9x & 2.40M & 59.4 & 5.43x \\
Email-Eu & $M_3$ & 12.0K & 40.1 & 493 & 1.15 & 34.7x & 276K & 16.8 & 6.22x \\
Email-Eu & $M_4$ & 554 & 4.99 & 23 & 1.09 & 4.57x & 49.2K & 6.44 & 0.775x \\
Email-Eu & $M_5$ & 57.7M & 7.18K & 4.71M & 131 & 54.9x & 144M & 1.75K & 4.10x \\
Email-Eu & $M_6$ & 7.08M & 1.54K & 31.1K & 2.45 & 627x & 6.89M & 200 & 7.68x
\\
\hline
\hline
& & \textbf{Static} & & \textbf{$\delta$ = 1 day} & & & \textbf{$\delta$ = 1 week} & & \\
  \textbf{Graph} & \textbf{Motif} & \textbf{Subgraphs} & \textbf{Time} & \textbf{Subgraphs} & \textbf{Time} & \textbf{Speed Up} & \textbf{Subgraphs} & \textbf{Time} & \textbf{Speed Up} \\
\hline
MathOverflow & $M_1$ & 1.42M & 2.94K & 2 & 0.001 & 2.94Mx & 531 & 0.047 & 62.5Kx \\
MathOverflow & $M_2$ & 218M & 11.4K & 248 & 0.015 & 760Kx & 9.06K & 0.218 & 52.3Kx \\
MathOverflow & $M_3$ & 5.53M & 4.02K & 4 & 0.001 & 4.02Mx & 684 & 0.047 & 85.5Kx \\
MathOverflow & $M_4$ & 285K & 2.34K & 0 & 0.015 & 156Kx & 180 & 0.015 & 156Kx \\
MathOverflow & $M_5$ & --- & $> 24$ hr & 805 & 0.031 & $>$ 2.79Mx & 214K & 5.897 & 14.7Kx \\
MathOverflow & $M_6$ & --- & $> 24$ hr & 1.70K & 0.062 & $>$ 1.39Mx & 72.0K & 2.059 & 42.0Kx \\
\hline
Enron & $M_1$ & 246K & 1.21K & 2 & 0.046 & 26.3Kx & 125 & 0.156 & 7.77Kx \\
Enron & $M_2$ & 119M & 4.51K & 1.21K & 0.078 & 57.9Kx & 22.7K & 0.656 & 6.88Kx \\
Enron & $M_3$ & 1.71M & 1.62K & 2 & 0.047 & 34.4Kx & 274 & 0.156 & 10.4Kx \\
Enron & $M_4$ & 33.3 K & 55.6 & 0 & 0.047 & 1.18Kx & 14 & 0.109 & 510x \\
Enron & $M_5$ & --- & $> 24$ hr & 12.6M & 333 & $>$ 259x & 217M & 6.05K & $>$ 14.3x \\
Enron & $M_6$ & --- & $> 24$ hr & 2.33K & 0.28 & $>$ 309Kx & 112K & 5.57 & $>$ 15.5Kx \\
\hline
\end{tabular}
\end{small}
\caption{Performance comparing the results of static subgraph matching versus our $\delta$-temporal matching for our test datasets. Different values of $\delta$ were used depending on the frequency of interactions in each network.  All times are in wall-clock seconds.}
\end{table*}

The motifs used can be seen in Fig 4.  These represent six different temporal subgraph patterns, between 4 and 5 nodes in size.  These were selected in part as they cannot currently be found by the more efficient algorithms of Paranjape et al., which only work for 2-node, triangle and 3-edge star motif patterns \cite{paranjape2017motifs}.  We chose query graphs which we believed would give a wide range of results on our test datasets.  We performed our experiments using $\delta =$ 1 hour and $\delta =$ 1 day for CollegeMSG and Email-Eu datasets, and $\delta =$ 1 day and $\delta =$ 1 week for the MathOverflow and Enron datasets.  The latter was necessary due to the longer periods of time between interactions in these networks.

To give an idea of the efficiency of our approach over that of using a static subgraph matching algorithm, we have calculated the following for each of our motifs:

\begin{itemize}
	\item The number of matching static subgraphs.
	\item The time in seconds to find the matching static subgraphs with the Boost implementation of VF2. \cite{siek2002boost}	
	\item The number of matching temporal subgraphs.
	\item The time in seconds to find the matching temporal subgraphs with our chronological edge-driven algorithm.
\end{itemize}

All algorithms were implemented in C++ and executed on a single thread of a 1.80 GHz Intel Xeon E5-2603 CPU with 64 GB of memory.  The Boost graph library implementation of the VF2 algorithm was used to perform the static subgraph matching in our experiments \cite{siek2002boost}.    Timings include only the time to perform subgraph searches, and not input or output times.  The results of our experiments can be seen in Table 3.  In some cases the time taken for VF2 to complete took over 24 hours.  In these situations we stopped the algorithm, and simply recorded the time as taking $>$ 24 hours.

Both the subgraph counts and performance times lead to some intriguing results.  In some cases, the difference between the number of static subgraphs and $\delta$-temporal subgraphs can be very dramatic.  The most extreme case was that of the MathOverflow graph, which had around 285,000 static subgraphs matching $M_4$, but none matching that pattern when time was taken into account for $\delta$ values of 1 day.  

In other cases, the number of $\delta$-temporal subgraphs was actually higher than the number of static subgraphs.  This is due to the fact that we were merged parallel edges when computing static subgraph isomorphism with VF2.  In some of these cases our approach was still significantly faster.  Examples include the CollegeMSG graph, when $\delta =$ 1 day, for motifs $M_1$, $M_2$, $M_3$, $M_4$, and $M_6$.  In each case, the number of temporal isomorphic subgraphs was significantly higher than the number of static isomorphic subgraphs, and yet there was still speed up between 1.26x and 4.66x.  We see a similar pattern with the Email-Eu datasets with $\delta = 1$ day as well.  In all but one case, our approach was faster than that of VF2, despite having more isomorphic subgraphs.

Generally speaking, increasing $\delta$ does decrease the performance of our approach.  In fact, when searching for the $M_4$ motif with $\delta = 1$ day in the Email-Eu graph, we actually found our algorithm to be slower than VF2.  We suspect this would continue to be the case for many of the datasets as $\delta$ continues to increase, although it can depend on many factors about the graph, such as the number of parallel edges between nodes, and the range of time represented in the network.

For the most part, the performance of our algorithm was vastly faster than that of VF2 for the given motifs.  This was especially true for the largest of the graphs, the MathOverflow and Enron datasets.  For these graphs, when $\delta =$ 1 day, we had speed up between 1000 and over 4 million times that of the VF2 algorithm, with 4 of the motifs for the MathOverflow producing over 1 million times speed up.  While the speed up decreased when $\delta$ = 1 week, we still saw speed ups between around 14x to over 100,000x that of VF2, with most motifs giving a speed up in the thousands.

To gain a better understanding of how the number of matching subgraphs effects the speed up of the algorithm, we created a chart plotting the ratio of static subgraphs over temporal subgraphs on the x-axis, with the amount of speed up on the y-axis.  The resulting log-log plot can be seen in Fig 5.    While there is some fluctuation in the results, there is a clear positive correlation between the ratio of matching subgraphs and the speed up of the algorithm.  While there are undoubtedly a number of factors that impact the overall performance, the reduction in number of matching subgraphs seems to be one of the biggest factors in our algorithm's performance improvement over using VF2.

\section{Conclusion and Future Work}

As can be seen in from the results of our experiments, a significant speed up can be had in many cases by applying a chronological edge-based matching approach to the temporal subgraph isomorphism problem.  Not surprisingly, the data also shows that the choice of $\delta$ can have a significant impact on the performance as well.  In situations where $\delta$ grows large, other techniques for temporal subgraph matching may prove to be more efficient.  This ultimately seems to depend heavily on the total number of matching subgraphs.  If the number of temporal isomorphic subgraphs greatly exceeds the number of static isomorphic subgraphs, it may end up being more efficient to follow the techniques that apply the static subgraph matching first, and filter for temporal matches second.  Further study will be needed to see if the secondary filtering technique of Paranjape et al \cite{paranjape2017motifs} is efficient enough to outperform our algorithm in these situations.

\begin{figure}
\includegraphics[scale=.32]{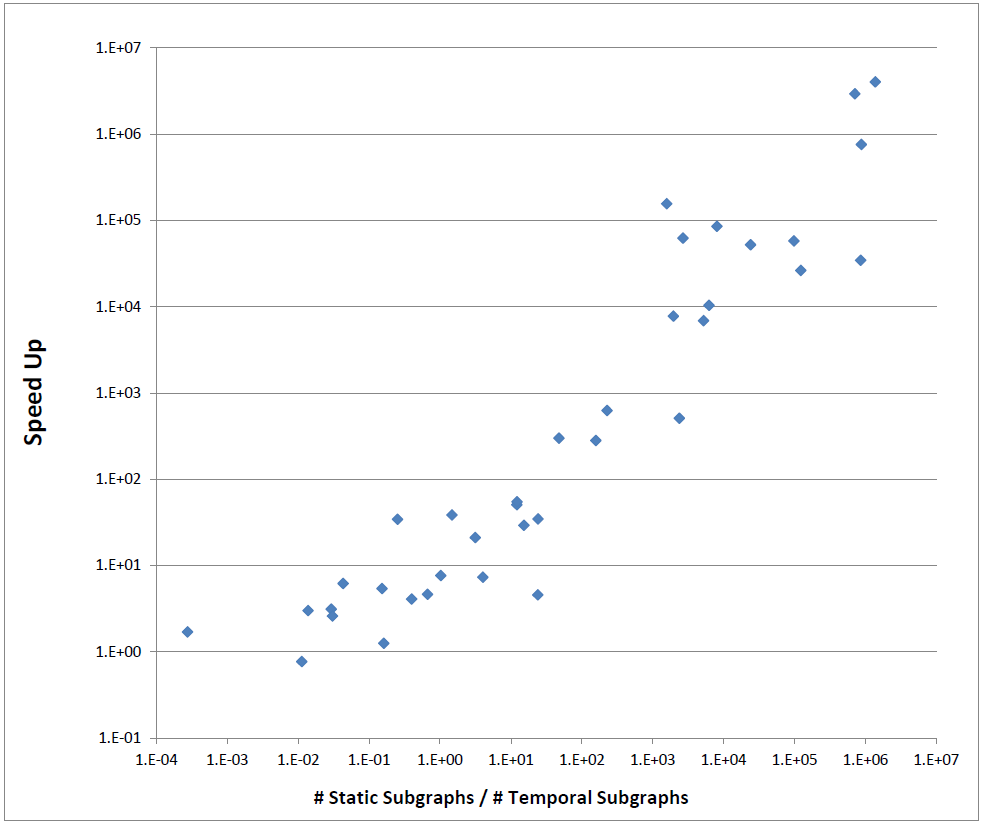}
\caption{A log-log plot comparing the ratio of static and temporal subgraphs to the overall speed up of our algorithm.}
\end{figure}

One of the fascinating differences between temporal and static subgraph matching is that the number of matching temporal graphs can be either greater or lesser than the number of static graphs.  Despite the fact that a number of the experiments produced \textit{more} matching subgraphs for the temporal approach, the performance was still usually faster than the traditional static matching method. It was only when the difference became two orders of magnitude greater that we saw a slow down of our approach over that of VF2.  We believe this advantage of our algorithm may in part be due to the fact that less of the search space is required to be traversed in the temporal matching process, even if it means more subgraphs are matched overall.  By only visiting edges in chronological order, we can often ignore a large portion of the edges during the matching process, improving the overall time complexity.

While smaller query graphs may suffice for many uses of motif analysis, we believe the ability of our approach to generalize to larger subgraphs could be useful to many.  It also shows that a more efficient general approach to the temporal subgraph matching problem is possible.  By performing the matching against the edges and not the nodes, we can take the ordering of the edges into account and match only those that are in the correct chronological sequence.  By making these algorithmic improvements, we can enable a speed up of performance by the thousands or even millions in some cases.

Subgraph matching and motif analysis have become an increasingly powerful tools for those working in data mining and graph analytics.  While the problem remains NP-complete, great gains can still be made in practice for many real-world datasets.  When it comes to query graphs of four or more nodes, we believe our approach to general temporal subgraph matching should be more efficient in most cases than previously published techniques.  By taking advantage of the chronological ordering of edges, larger temporal motifs and query graphs should no longer be off limits from being used in temporal network analysis.

\section{Acknowledgements}

The research described is part of the Analysis in Motion (AIM) Initiative at Pacific Northwest National Laboratory (PNNL). It was conducted under the Laboratory Directed Research and Development Program at PNNL, a multi-program national laboratory operated by Battelle for the U.S. Department of Energy under Contract DE-AC05-76RL01830. 

\bibliographystyle{siam}
\bibliography{temporal_subgraph_matching}

\end{document}